%% file: ParsaiFSEN.tex
\documentclass[runningheads,a4paper]{llncs}
\usepackage{changes}
\usepackage{amssymb}
\usepackage{graphicx}
\usepackage{url}
\usepackage[boxed]{algorithm2e}

\usepackage{graphicx}
\usepackage[normalem]{ulem}
\usepackage{adjustbox}
\usepackage{hhline}
\usepackage{multirow}
\usepackage{url}
\usepackage{rotating}
\usepackage[cmex10]{amsmath}
\usepackage{balance}
\usepackage[nocompress]{cite}

\newcommand{\keywords}[1]{\par\addvspace\baselineskip
\noindent\keywordname\enspace\ignorespaces#1}

\begin{document}

\mainmatter

\title{LittleDarwin: a Feature-Rich and Extensible\\ Mutation Testing Framework for \\Large and Complex Java Systems}

\titlerunning{LittleDarwin: An Extensible Mutation Testing Framework for Java Systems}

\author{Ali Parsai\and Alessandro Murgia\and Serge Demeyer}

\authorrunning{Ali Parsai\and Alessandro Murgia\and Serge Demeyer}

\institute{Antwerp Systems and Software Modelling Lab\\
    University of Antwerp\\
    \url{{ali.parsai, alessandro.murgia, serge.demeyer}@uantwerpen.be}
    }

\maketitle

\begin{abstract}
Mutation testing is a well-studied method for increasing the quality of a test suite. We designed LittleDarwin as a mutation testing framework able to cope with large and complex Java software systems, while still being easily extensible with new experimental components. LittleDarwin addresses two existing problems in the domain of mutation testing: having a tool able to work within an industrial setting, and yet, be open to extension for cutting edge techniques provided by academia. LittleDarwin already offers higher-order mutation, null type mutants, mutant sampling, manual mutation, and  mutant subsumption analysis. There is no tool today available with all these features that is able to work with typical industrial software systems.

\end{abstract}

\input{1-intro}

\input{2-background}
\input{3-littledarwin}

\input{4-experiments}

\input{9-conclusion}

\subsubsection*{Acknowledgments}
This work is sponsored by the Institute for the Promotion of Innovation through Science and Technology in Flanders through a project entitled Change-centric Quality Assurance (CHAQ) with number 120028.

\balance

\bibliographystyle{splncs03}
\bibliography{Master}

\end{document}

%% file: 1-intro.tex
\section{Introduction}
Along with the popularity of agile methods in recent times came an emphasis on test-driven development and continuous integration~\cite{Beck2003,Fowler2006}. This implies that developers are interested in testing their software components early and often~\cite{McGregor2007}. 
Therefore, the quality of the test suite is an important factor during the evolution of the software. One of the extensively studied methods to improve the quality of a test suite is mutation testing~\cite{DeMillo1978}.  

Mutation testing was first proposed by DeMillo, Lipton, and Sayward to measure the quality of a test suite by assessing its fault detection capabilities~\cite{DeMillo1978}.
Mutation testing has been shown to simulate faults realistically~\cite{Andrews2005,Just2014}. This is because the faults introduced by each mutant are modeled after  common mistakes developers  make~\cite{Jia2011}. Mutation testing is demonstrated to be  a more powerful coverage criteria in comparison with data-flow, statement, and branch coverage~\cite{Walsh1985,Frankl1997}. 
 
Recent trends in scientific literature indicate a surge in popularity of this technique, along with an increased usage of real projects as the subjects of scientific experiments~\cite{Jia2011}. In literature, topics such as creating more robust mutants using higher-order mutation~\cite{Jia2009,Papadakis2010,Kintis2012,Omar2014}, reducing redundancy among mutants using mutant subsumption~\cite{Ammann2014,Papadakis2016,Kurtz2016}, and reducing the number of mutants using mutant selection~\cite{Zhang2013,Gligoric2013a,Gopinath2015} are gaining popularity. 
Despite its benefits, the idea of mutation testing  is not widely used in industry. Consequently, mutation testing research   stays behind 
since it lacks fundamental experiments on industrial software systems.
We believe that, beyond the computationally expensive nature of mutation testing~\cite{Offutt2001},
the reluctance of industry can  stem from the shortage of mutation testing tools that can both (i) work on large and complex systems, and (ii) incorporate new and upcoming techniques as an experimental framework. 

In this paper, we try to fill this gap by introducing LittleDarwin. LittleDarwin is designed as a mutation testing framework aiming to target large and complex systems. The design decisions are geared towards a simple architecture that allows the addition of new experimental components, and fast prototyping. 
In its current version, LittleDarwin facilitates experimentation on higher-order mutation, null type mutants, mutant sampling, manual mutation, and mutant subsumption analysis. LittleDarwin has been used for experimentation on several large and complex open source and industrial projects~\cite{Parsai2015T,Parsai2016,Parsai2016M}.  %

The rest of the paper is structured as follows. 
We provide background information about mutation testing in Section~\ref{section:Background}. 
We explain the design and the implementation of our tool in Section~\ref{section:Design},
and summarize the experiments that have been performed using our tool in Section~\ref{section:Experiments}. 
We conclude the paper in Section~\ref{section:Conclusion}.

%% file: 2-background.tex
\section{Mutation Testing}
\label{section:Background}
Mutation testing\footnote{The idea of mutation testing was first mentioned by Lipton, and later developed by DeMillo, Lipton and Sayward~\cite{DeMillo1978}. The first implementation of a mutation testing tool was done by Timothy Budd in 1980~\cite{Budd1980}.} is the process of injecting faults into a software system  to verify whether the test suite detects the injected fault.
Mutation testing starts with a \textit{green} test suite --- a test suite in which all the tests pass. First, a faulty version of the software is created by introducing faults into the system \textit{(Mutation)}. This is done by applying a known transformation \textit{(Mutation Operator)} on a certain part of the code.  After generating the faulty version of the software \textit{(Mutant)}, it is passed onto the test suite. If there is an error or failure during the execution of the test suite, the mutant is marked as killed \textit{(Killed Mutant)}. If all tests pass, it means that the test suite could not catch the fault, and the mutant has survived \textit{(Survived Mutant)}~\cite{Jia2011}.

\textbf{Mutation Operators.} A mutation operator is a transformation which introduces a single syntactic change into its input. The first set of mutation operators were reported in King et al.~\cite{King1991}. These mutation operators work on essential syntactic entities of the programming language such as arithmetic, logical, and relational operators. They were introduced in the tool Mothra which was designed to mutate the programming language FORTRAN77. In 1996, Offutt et al. determined that a selection of few mutation operators is enough to produce similarly capable test suites  with a four-fold reduction of the number of mutants~\cite{Offutt1996}. This reduced-set of operators
 remained more or less intact in all subsequent research papers. 
With the advent of object-oriented programming languages, new mutation operators were proposed to cope with the specifics of this programming paradigm~\cite{Kim2000,Ma2002}. %

\textbf{Equivalent Mutants.} 
If the output of a mutant for all possible input values is the same as the original program, it is called an \emph{equivalent mutant}. It is not possible to create a test case that passes for the original program and fails for an equivalent mutant, because the equivalent mutant is indistinguishable from  the original program. This makes the  creation of equivalent mutants undesirable, and leads to false positives during mutation testing.  In general, detection of equivalent mutants is undecidable due to the halting problem~\cite{Offutt1997}. Manual inspection of all mutants is the only way of filtering all equivalent mutants, which is impractical in real projects due to the amount of work it requires. Therefore, the common practice within today's state-of-the-art is to take precautions to generate as few equivalent mutants as possible, and accept equivalent mutants as a threat to validity (accepting a false positive is less costly than removing a true positive by mistake~\cite{Fawcett2006}). 

\textbf{Mutation Coverage.} 
Mutation testing allows software engineers to monitor the fault detection capability of a test suite by means of mutation coverage (see Equation~\ref{coverageequation})~\cite{Jia2011}.
A test suite is said to achieve \textit{full mutation test adequacy} whenever it can kill all the non-equivalent mutants, thus reaching a mutation coverage of 100\%. Such test suite is called a \textit{mutation-adequate test suite}. 

\begin{equation}
		Mutation\ Coverage = \frac{Number\ of\ killed\ mutants}{Number\ of\ all\ non\mbox{-}equivalent\ mutants}
	\label{coverageequation}
\end{equation}

\textbf{Higher-Order Mutants.} 
First-order mutants are the mutants generated by applying a mutation operator on the source code only once.  
By applying mutation operators more than once  we obtain higher-order mutants. Higher-order mutants can also be described as a combination of several first-order mutants. 
Jia et al. introduced the concept of higher-order mutation testing and discussed the relation between higher-order mutants and first-order mutants~\cite{Jia2008}. 

\textbf{Mutant Subsumption.}
Mutant subsumption is defined as the relationship between two mutants \texttt{A} and \texttt{B} in which \texttt{A} subsumes \texttt{B} if and only if the set of inputs that kill \texttt{A} is guaranteed to kill \texttt{B}~\cite{Kurtz2015}. The subsumption relationship for faults has been defined by Kuhn in 1999~\cite{Kuhn1999}, but its use for mutation testing has been popularized by Jia et al. for creating hard to kill higher-order mutants~\cite{Jia2008}. Later on, Ammann et al.  tackled the theoretical side of mutant subsumption~\cite{Ammann2014}. In their paper, Ammann et al. define \textit{dynamic} mutant subsumption, which redefines the relationship using test cases. Mutant \texttt{A} dynamically subsumes Mutant \texttt{B} if and only if (i) \texttt{A} is killed, and (ii) every test that kills \texttt{A} also kills \texttt{B}.
The main purpose behind the use of mutant subsumption is to reliably detect redundant mutants, which create multiple threats to the validity of mutation testing~\cite{Papadakis2016}. This is often done by determining the dynamic subsumption relationship among a set of mutants, and keeping only those that are not subsumed by any other mutant.

\textbf{Mutant Sampling.}
To make mutation testing practical, it is important to reduce its execution time. One way to achieve this is to reduce the number of mutants. A simple approach to mutant reduction is to randomly select a set of mutants. This idea was first proposed by Acree~\cite{Acree1980} and Budd~\cite{Budd1980} in their PhD theses. 
To perform random mutant sampling, no  extra information regarding the context of the mutants is needed. This makes the implementation of this technique in  mutation testing tools easier. Because of this, and the simplicity of  random mutant sampling, its performance overhead  is negligible. 
Random mutant sampling can be performed uniformly, meaning that each mutant has the same chance of being selected.
Otherwise, random mutant sampling can be enhanced  by using heuristics based on  the source code. 
The percentage of mutants that are selected determines the \textit{sampling rate} for random mutant sampling. %

%% file: 3-littledarwin.tex
\section{Design and Implementation}
\label{section:Design}
In this section, we discuss the implementation details of LittleDarwin, and provide information  on our design decisions.

\subsection{Algorithm}
LittleDarwin is designed with simplicity in mind, in order to increase the flexibility of the tool.  %
To this effect, it mutates the Java source code rather than the byte code in order to defer the responsibility of compiling and executing the code to the build system. This allows LittleDarwin  to remain as flexible as possible regarding the complexities stemming from the build and test structures of the target software.  
The procedure is divided into two phases: \textit{Mutation Phase} (Algorithm~\ref{alg:mutationphase}), and \textit{Test Execution Phase} (Algorithm~\ref{alg:testexecphase}). 

\textbf{Mutation Phase.}
In this phase, the tool creates the mutants for each source file. LittleDarwin first searches for all source files contained in the path given as input, and adds them to the processing queue. Then, it selects an unprocessed source file from the queue, parses it, applies all the mutation operators, and saves all the generated mutants.  

\begin{algorithm}[H]
    \caption{Mutation Phase}
    \label{alg:mutationphase}
    \SetKwData{Queue}{queue} \SetKwData{Src}{srcFile} \SetKwArray{Mutated}{mutants}
    \SetKwFunction{Mutate}{mutate}
    \SetKwInOut{Input}{Input}\SetKwInOut{Output}{Output}
    
    \Input{Java source files}
    \Output{Mutated Java source files}

    \BlankLine   
    \Queue $\leftarrow$ all Java source files\;
    \While{\Queue $\neq \emptyset$}{
        \Src $\leftarrow$ \Queue .pop()\;
        \Mutated{\Src} $\leftarrow$  \Mutate{\Src}\;
        
    }
    \Return \Mutated\;      
    \BlankLine

\end{algorithm}

\textbf{Test Execution Phase.} 
In this phase, the tool executes the test suite for each mutant. First the build system is executed without any change to ensure that the test suite runs ``green''. Then, a source file along with its mutants are read from the database, and the output of the build system is recorded for each mutant. If the build system fails (exits with non-zero status) or times out, the mutant is categorized as killed. If the build system is successful (exits with zero status), the mutant is categorized as survived. Finally, a report is generated for each source file, and an overall report is generated for the project (see Figure~\ref{fig:report-screenshot} for an example of this).

\begin{algorithm}[H]
    
    \caption{Test Execution Phase}
    \label{alg:testexecphase}
    \SetKwData{Queue}{queue} \SetKwData{Src}{srcFile} \SetKwData{Mut}{mutantFile} 
    \SetKwArray{Mutated}{mutants}
    \SetKwArray{Results}{result}
    
    \SetKwFunction{Mutate}{mutate}
    \SetKwFunction{Replace}{replace}
    \SetKwFunction{Backup}{backup}
    \SetKwFunction{Restore}{restore}

    \SetKwFunction{Build}{executeTestSuite}
    \SetKwInOut{Input}{Input}\SetKwInOut{Output}{Output}
    \Input{Mutated Java source files}
    \Output{Mutation Testing Report}
    \BlankLine   
    
    \If{\Build{} is successful}
    {
        \ForEach{\Src}{\Queue $\leftarrow$   \Mutated{\Src}\;
            \Backup{\Src}\;
            \While{\Queue $\neq \emptyset$}{
                \Mut $\leftarrow$ \Queue .pop()\;
                \Replace{\Src,\Mut}\;
                \Results{\Mut} $\leftarrow$ \Build{}\;
            }
            \Restore{\Src}\;
            Generate report for \Src\;
            
        }
        Generate overall report\;
    }
    \Return reports\;      
    \BlankLine  
    
\end{algorithm}

\subsection{Components}
The data flow diagram of the main internal components of LittleDarwin is shown in Figure~\ref{fig:littledarwincomponents}. The following is an explanation of each main component:

\begin{figure}
    \centering
    \includegraphics[width=0.8\linewidth]{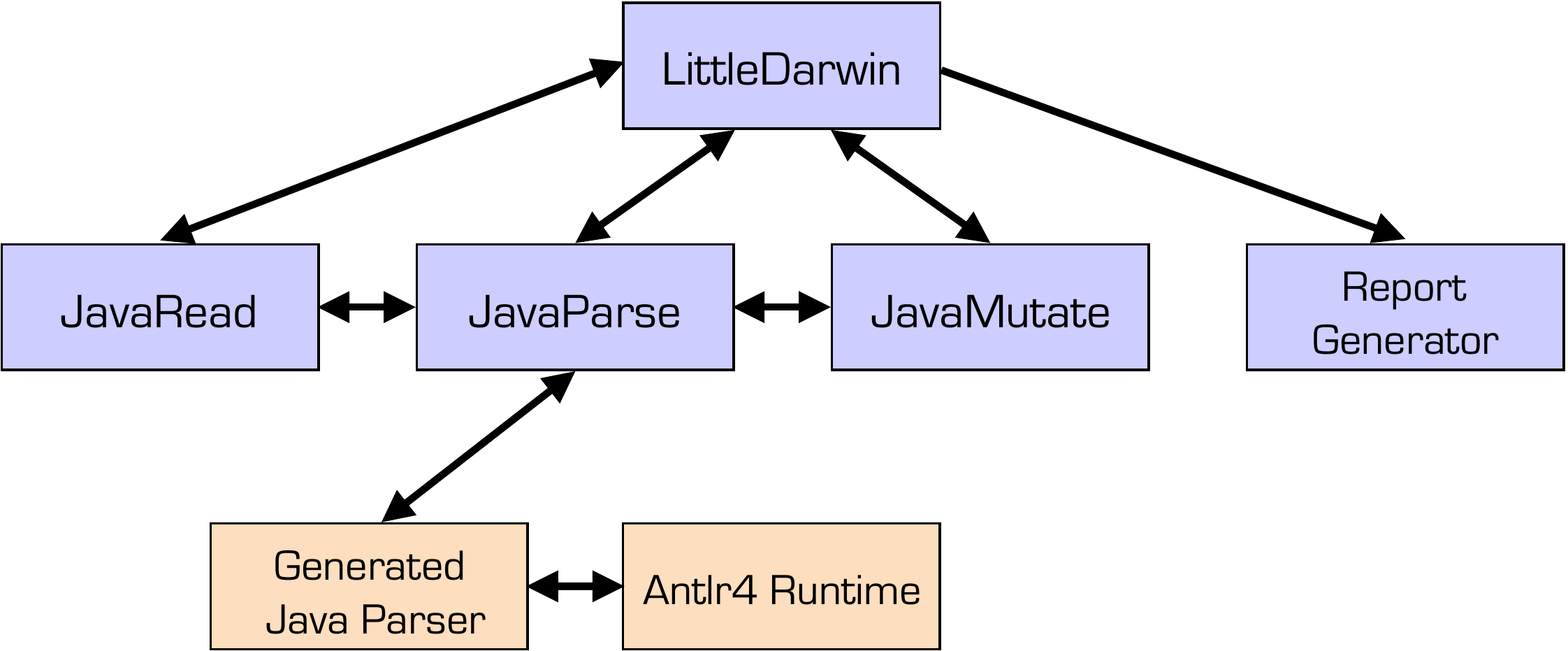}
    \caption{Data Flow Diagram for LittleDarwin Components}
    \label{fig:littledarwincomponents}
\end{figure}

\textbf{JavaRead.} This component provides methods to perform input/output operations on Java files. LittleDarwin uses this component to read the source files, and write the mutants back to disk.

\textbf{JavaParse.} This component parses Java files into an abstract syntax tree. This is necessary to produce valid and compilable  mutants. To implement this functionality, an Antlr4\footnote{\url{http://www.antlr.org/}} Java 8 grammar is used along with a customized version of Antlr4 runtime. Beside providing the parser, this component also provides the functionality to pretty print the modified tree back to a Java file. 

\begin{figure}
	\centering
	\fbox{ 	\includegraphics[width=0.6\linewidth]{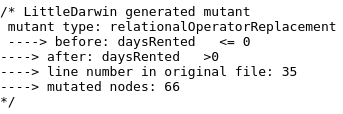}}
	\caption{The Header of a LittleDarwin Mutant}
	\label{fig:mutantheader}
\end{figure}

 \textbf{JavaMutate.} This component manipulates the abstract syntax tree (AST) created by the parser. Subsection~\ref{sec:mutationoperators} explains the mutation operators of LittleDarwin in detail. The currently implemented mutation operators search the provided AST for mutable nodes matching the predefined patterns (for example, \textit{AOR-B} looks for all binary arithmetic operator nodes that do not contain a string as an operand), and they perform the mutation on the tree itself. This gives the developer flexibility in creating new complicated mutation operators. Even if a mutation operator introduces a fault that needs to change several statements at once, and depends on the context of the statements, it can be implemented using a complicated search pattern on the AST. 
The mutation operators are designed to exclude mutations that would lead to compilation errors. However, not all of these cases can be detected using an AST (e.g. AOR-B on two variables that contain strings). Handling of such cases are therefore left for the post-processing unit that filters such mutants based on the output of the Java compiler.
In order to preserve the maximum amount of information for post-processing purposes, for each mutant a  commented header is created. This header contains the following information: (i) the mutation operator that created the mutant, (ii) the mutated statement before and after the mutation, (iii) the line number of the mutated statement in the original source file, and (iv) the id number of the mutated node(s). An example  is shown in Figure~\ref{fig:mutantheader}.

\begin{figure}
 	\centering
\fbox{ 	\includegraphics[width=0.65\linewidth]{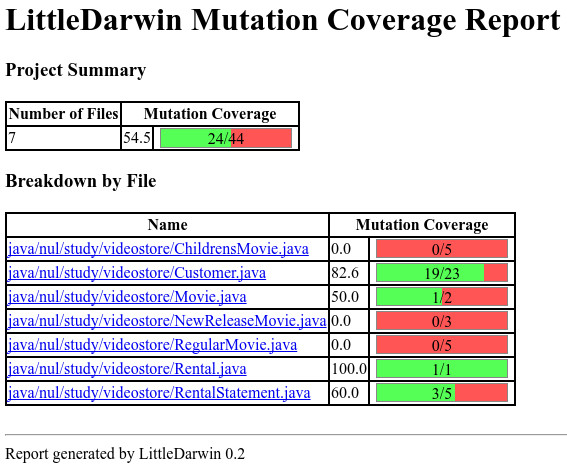}}
 	\caption{LittleDarwin Project Report}
 	\label{fig:report-screenshot}
 \end{figure}

 \textbf{Report Generator.} This component generates HTML reports for each file. These reports contain all the generated mutants and the output of the build system after the execution of each mutant.  In the end, an overall report is generated for the whole project (Figure~\ref{fig:report-screenshot}).

\subsection{Mutation Operators of LittleDarwin}
\label{sec:mutationoperators}
There are 9 default mutation operators implemented in LittleDarwin listed in Table~\ref{mutationoperators}.
These operators are based on the reduced-set of mutation operators that were demonstrated by Offutt et al. to be capable of creating similar-strength test suites as the full set of mutation operators~\cite{Offutt1996}.
Since the number of mutation operators of LittleDarwin is limited, it is possible that no mutants are generated for a class that lacks mutable statements. In practice, we observed that usually only very small compilation units (e.g. interfaces, and abstract classes) are subject to this condition.  

\begin{table}[!h]
	\footnotesize
	\centering
	
	\caption{LittleDarwin Mutation Operators}
	\label{mutationoperators}
	
	\begin{tabular}{|l||l|c|c|}
		\hline \multirow{2}{*}{\textbf{Operator}} & \multirow{2}{*}{\textbf{Description}} & \multicolumn{2}{c|}{\textbf{Example}} \\
		\hhline{~~--} & & \textbf{Before} & \textbf{After} \\ 
		\hline
		\hline AOR-B & Replaces a binary arithmetic operator & $a + b$  & $a - b$ \\ 
		\hline AOR-S & Replaces a shortcut arithmetic operator & $++a$ & $--a$ \\ 
		\hline AOR-U & Replaces a unary arithmetic operator & $-a$ & $+a$ \\ 
		\hline LOR & Replaces a logical operator & $a\,\&\,b$ & $a\,|\,b$ \\ 
		\hline SOR & Replaces a shift operator & $a >> b$ & $a << b$ \\ 
		\hline ROR & Replaces a relational operator & $a >= b$ & $a < b$ \\ 
		\hline COR & Replaces a binary conditional operator & $a\:\&\&\:b$ & $a\,||\,b$ \\ 
		\hline COD & Removes a unary conditional operator & $!\,a$  & $a$ \\ 
		\hline SAOR & Replaces a shortcut assignment operator & $a\:*= b$ & $a\:/= b$ \\ 
		\hline 
	\end{tabular}
	\end{table}

In addition to these mutation operators, there are four experimental mutation operators in LittleDarwin that are designed to simulate null type faults. These mutation operators along with the faults they simulate are provided in  Table~\ref{table:nulloperators}. We included these mutation operators based on the conclusions offered by Osman et al.~\cite{Osman2014}. In their study, they discover that the null object is a major source of software faults. The null type mutation operators are able to simulate such faults, and consequently assess the quality of the test suite with respect to them. These mutation operators cover fault-prone aspects of a method: \textit{NullifyInputVariable} mutates the method input, \textit{NullifyReturnValue} mutates the method output, and \textit{NullifyObjectInitialization} and  \textit{RemoveNullCheck} mutate the statements in method body.

\begin{table}
	\centering
	\caption{Null Type Faults and Their Corresponding Mutation Operators}    
	\label{table:nulloperators}
\adjustbox{max width=\linewidth}{
	\begin{tabular}{|c|c|c|}
        \hline
        \textbf{Fault}                                                                   & \textbf{Mutation Operator}  & \textbf{Description}                                                                                                         \\ \hline \hline
        \begin{tabular}[c]{@{}c@{}}Null is returned\\ by a method\end{tabular}           & NullifyReturnValue          & \begin{tabular}[c]{@{}c@{}}If a method returns an object,\\  it is replaced by \texttt{null}\end{tabular}                  \\ \hline
        \begin{tabular}[c]{@{}c@{}}Null is provided \\ as input to a method\end{tabular} & NullifyInputVariable        & \begin{tabular}[c]{@{}c@{}}If a method receives an object \\reference,  it is replaced by \texttt{null}\end{tabular}       \\ \hline
        \begin{tabular}[c]{@{}c@{}}Null is used to \\ initialize a variable\end{tabular} & NullifyObjectInitialization & \begin{tabular}[c]{@{}c@{}}Wherever there is a \texttt{new} statement, \\ it is replaced with \texttt{null}\end{tabular} \\ \hline
        \begin{tabular}[c]{@{}c@{}}A null check\\  is missing\end{tabular}               & RemoveNullCheck             & \begin{tabular}[c]{@{}c@{}}Any binary relational statement \\ containing \texttt{null} at one side is negated\end{tabular} \\ \hline
    \end{tabular}
}
\end{table}

\subsection{Design Characteristics}
To foster mutation testing in industrial setting it is important to have a tool able to work on large and complex systems.
Moreover, to allow researchers to use real-life projects as the subjects of their studies, it is also important to  provide a framework that is easy to extend.
In this section, we show to what extent LittleDarwin, and its main alternatives, can satisfy these requirements. 
As alternatives, we use PITest~\cite{Coles2016},  Javalanche~\cite{Schuler2009a}, and MuJava~\cite{Ma2006a},
since they are popular tools used in literature. In Table~\ref{table:mutationtools}, we summarize the design highlights.

\begin{table}[]
\centering
\caption{Comparison of Features in Mutation Testing Tools}
\label{table:mutationtools}
    \adjustbox{max width=\linewidth}{
\begin{tabular}{|l|l|c|c|c|c|}
\hline
\multicolumn{2}{|l|}{\textbf{Features}} & \textbf{LittleDarwin} & \textbf{PITest}~\cite{PITest} & \textbf{Javalanche}~\cite{Schuler2009a} & \textbf{MuJava}~\cite{Ma2005} \\ \hline \hline
\multirow{4}{*}{Compatibility with} & Maven & $\checkmark$ & $\checkmark$ & $\times$ & $\times$ \\ \cline{2-6}
 & Ant & $\checkmark$ & $\checkmark$ & $\times$ & $\times$ \\ \cline{2-6}
 & Gradle & $\checkmark$ & $\checkmark$ & $\times$ & $\times$ \\ \cline{2-6}
 & Others & $\checkmark$ & $\times$ & $\times$ & $\times$ \\ \hline
\multicolumn{2}{|l|}{Support for Complex Test Structures} & $\checkmark$ & $\times$ & $\times$ & $\times$ \\ \hline
\multicolumn{2}{|l|}{Optimized for Performance} & $\times$ & $\checkmark$ & $\checkmark$ & $\checkmark$ \\ \hline
\multicolumn{2}{|l|}{Optimized for Experimentation} & $\checkmark$ & $\times$ & $\times$ & $\times$ \\ \hline
\multicolumn{2}{|l|}{Tested on Large Systems} & $\checkmark$ & $\checkmark$ & $\checkmark$ & $\times$ \\ \hline
\multicolumn{2}{|l|}{Ability to Retain Detailed Results} & $\checkmark$ & $\times$ & $\times$ & $\checkmark$ \\ \hline
\multicolumn{2}{|l|}{Open Source} & $\checkmark$ & $\checkmark$ & $\checkmark$ & $\checkmark$ \\ \hline
\end{tabular} }
\end{table}

\textbf{Compatibility with Major Build Systems.}
To make the initial setup of  a mutation testing tool easier, it needs to work with popular build systems for Java programs. 
LittleDarwin executes the build system rather than integrate into it, and therefore, can readily support various build systems. In fact, the only restrictions imposed by LittleDarwin are: (i) the build system must be able to run the test suite, and (ii) the build system must return non-zero if any tests fail, and zero if it succeeds. 
PITest address the challenge via integration into the popular build systems by means of plugins. 
At the time of writing it supports Maven\footnote{\url{https://maven.apache.org/}}, Ant\footnote{\url{https://ant.apache.org/}}, and Gradle\footnote{\url{https://gradle.org/}}. Javalanche and MuJava do not integrate in the build system.

\textbf{Support for Complex Test Structures.}
One of the difficulties of performing mutation testing on complex Java systems is to find and execute the test suite correctly. 
The great variety of testing strategies and unit test designs generally causes problems in executing the test suite correctly. %
LittleDarwin overcomes this problem thanks to a loose coupling with the test infrastructure, instead relying on the build system to execute the test suite. 
Other mutation testing tools reported in Table~\ref{table:mutationtools}  have problems in this regard.

\textbf{Optimized for Performance.} LittleDarwin mutates the source code and performs the execution of the test suite using the build system. 
This introduces a performance overhead for the analysis. For each mutant injected, LittleDarwin demands a rebuild and test cycle on the build system. 
The rest of the mutation tools use byte code mutation, which leads to better performance.

\textbf{Optimized for Experimentation.} 
LittleDarwin is written in Python to allow fast prototyping~\cite{Prechelt2000}. To parse the Java language, LittleDarwin uses an Antlr4 parser. This allows us to rapidly adapt to the syntactical changes in newer versions of Java (such as Java 8). This parser produces a complete abstract syntax tree that makes   the implementation of experimental features easier. In addition, the modular and multi-phase design of the tool allows reuse of each module independently. Therefore, it becomes easier to customize the tool according to the requirements of a new experiment.
The other mutation tools work on byte code, and therefore do not offer such facilities.

\textbf{Tested on Large Systems.}
LittleDarwin has been used in the past on software systems with more than 82 KLOC~\cite{Parsai2016,Parsai2016M}. 
PITest and Javalanche have been used in experiments with softwares of comparable size~\cite{Parsai2014,Schuler2009a}. 
We did not find  evidence that MuJava has been tested on large systems.

\textbf{Ability to Retain Detailed Results.} 
PITest and Javalanche  only output  a report on the killed and survived mutants. 
However, in many cases this is not enough. For example, subsumption analysis requires the name of all the tests that kill a certain mutant.
To address this problem, LittleDarwin retains all the output provided by the build system for each mutant, and allows for post-processing of the results. This also allows the researchers to manually verify the correctness of the results. %
MuJava provides an analysis framework as well, allowing for further experimentation~\cite{Ma2006a}.

\textbf{Open Source.}
LittleDarwin is a free and open source software system. The code of LittleDarwin and its components are provided\footnote{\url{https://github.com/aliparsai/LittleDarwin}} for  public use under the terms of GNU General Public License version 2. %
PITest and MuJava  are released under Apache License version 2. %
 Javalanche is released into public domain without an accompanying license.

\subsection{Experimental Features}
In order to facilitate the means for research in mutation testing, LittleDarwin supports several features up to date with the state of the art. A summary of these features and their availability in the alternative tools is provided in Table~\ref{table:mutationtoolsfeatures}. An explanation of  each feature follows.

\begin{table}[]
    \centering
    \caption{Comparison of Experimental Features in Mutation Testing Tools}
    \label{table:mutationtoolsfeatures}
    
    \begin{tabular}{|l|c|c|c|c|}
        \hline 
        \textbf{Experimental Features}& \textbf{LittleDarwin} & \textbf{PITest} & \textbf{Javalanche} & \textbf{MuJava} \\ 
        \hline  \hline 
        Higher-Order Mutation & $\checkmark$ & $\times$ & $\times$ & $\times$ \\ 
        \hline 
        Mutant Sampling & $\checkmark$ & $\times$ & $\times$ & $\checkmark$ \\ 
        \hline 
        Subsumption Analysis & $\checkmark$ & $\times$ & $\times$ & $\times$ \\ 
        \hline 
        Manual Mutation & $\checkmark$ & $\times$ & $\times$ & $\times$ \\ 
        \hline 
    \end{tabular} 
\end{table}

\textbf{Higher-order Mutation.}
This feature is designed to combine two first-order mutants into a higher-order mutant. It is possible to link the higher-order mutants to their first-order counterparts after acquiring the results. %

\textbf{Mutant Sampling.}
This feature is designed to use the results for sampling experiments. 
LittleDarwin by default implements two sampling strategies: uniform, and weighted. The uniform approach selects the mutants randomly with the same chance of selection for all mutants. 
In the weighted approach, a weight is assigned to each mutant that is proportional to the size of the class containing the mutant.
The given infrastructure also allows for the development of other techniques.

\textbf{Subsumption Analysis.}
This feature is designed to determine the subsumption relationship between mutants. For each mutant, this feature can determine whether the mutant is subsuming or not, which tests kill the mutant, which mutants are subsuming the mutant, and which mutants are subsumed by the mutant. It is also capable of exporting the mutant subsumption graph proposed by Kurtz et al.~for each project \cite{Kurtz2014,Kurtz2015}.

\textbf{Manual Mutation.}
This feature allows the researcher to use their manually created mutants with LittleDarwin. LittleDarwin is capable of automatically matching the mutants with the corresponding source files, and creating  the required structure to perform the analysis. For example, this is useful in case the mutants are  created with a separate tool.

%% file: 4-experiments.tex
\section{Experiments}
\label{section:Experiments}
In this section, we provide a brief summary of the experiments we already performed using the experimental features of LittleDarwin on large and complex  systems.

\textbf{Mutation Testing of a Large and Complex Software System.}
We used LittleDarwin to analyze a large and complex safety critical system for Agfa HealthCare.  Our attempts to use other mutation testing tools failed due to the complex testing structure of the target system. Due to this complexity, these tools were not able to detect the test suite. This is because (i) the project used OSGI\footnote{\url{https://www.osgi.org/developer/specifications/}} headers to dynamically load modules, and (ii) the test suite was located in a different component, and required several frameworks to work. 
The loose coupling of LittleDarwin with the testing structure allowed us to use the build system to execute the test suite, and thus, successfully perform mutation testing on the project.
For more details on this experiment, including the specification of the target system, and the run time of the experiment, please refer to Parsai's master's thesis~\cite{Parsai2015T}.

\textbf{Experimenting Up to Date Techniques on Real-Life Projects.}
LittleDarwin was used to perform three separate studies using the up to date techniques reported in Table \ref{table:mutationtoolsfeatures}. 
We were able to perform these studies on real-life projects.

In our study on random mutant sampling,  we noticed that related literature have two shortcomings~\cite{Parsai2016}. They focus their analysis at project level and they are mainly based on toy projects with adequate test suites. Therefore, we evaluated random mutant sampling at class level, and on real-life projects with non-adequate test suites. 
We used LittleDarwin to study two sampling strategies: uniform, and weighted. %
We highlighted that the weighted approach increases the chance of inclusion of mutants from classes with a small set of mutants in the sampled set, and reduces the viable sampling rate from 65\% to 47\% on average. 
This analysis was performed on 12 real-life open source projects.

In  our study on higher-order mutation testing, we used LittleDarwin to perform our experiments~\cite{Parsai2016M}.
We proposed a  model to estimate the first-order mutation coverage from higher-order mutation coverage. Based on this, we proposed a way  to halve the computational cost of acquiring  mutation coverage. In doing so, we achieved a strong correlation between the estimated and actual values. %
Since LittleDarwin retains the information necessary for post-processing the results, we were able to analyze the relationship between each higher-order mutant and its corresponding first-order mutants.  

We performed a study on simulating the null type faults which is currently under peer-review. 
In this study, we show that mutation testing tools are not adequate to strengthen the test suite against null type faults in practice. This is mainly because the traditional mutation operators of current mutation testing tools do not model null type faults. 
We implemented four new mutation operators in LittleDarwin to model null type faults explicitly, and we show how these mutation operators can be operatively used to extend the test suite in order to  prevent null type faults.  Using LittleDarwin, we were able to analyze the test suites of 15 real-life open source projects, and describe the trade offs related to the adoption of these operators to strengthen the test suite. 
We also used the mutant subsumption feature of LittleDarwin  to perform redundancy analysis on all 15 projects.

\textbf{Pilot Experiment.}
We performed a pilot experiment on a real life project in order to compare LittleDarwin with two of its alternatives: PITest and Javalanche. %
 In this experiment, we used Jaxen\footnote{\url{http://jaxen.org/}} as the subject, since it has been used before to evaluate Javalanche by its authors~\cite{Schuler2010}. Jaxen has 12,438 lines of production code, and 7,539 lines of test code. Table~\ref{table:pilot} shows the results of our pilot experiment. As we can see, even though LittleDarwin creates the least number of mutants, it is still slowest per-mutant. This is mainly because PITest and Javalanche both filter the mutants prior to analysis based on statement coverage. In addition, LittleDarwin relies on the build system to run the test suite, which introduces per-mutant overhead.

\begin{table}
    \centering
    \caption{Pilot Experiment Results}    
    \label{table:pilot}
    \adjustbox{max width=\linewidth}{
\begin{tabular}{|c|c|c|c|c|c|}
\hline    Tool & Generated Mutants & Killed Mutants & Mutation Coverage & Analysis Time & Per-mutant Time\\ 
\hline    LittleDarwin &  1,390 & 805 & 57.9\%  &  2h23m45s  & 6.21s\\ 
\hline    PITest & 4,315  & 2,145 & 49.8\% & 1h13m13s & 1.02s \\ 
\hline    Javalanche & 9,285  & 4,442 & 47.8\%  & 1h35m23s & 0.62s \\
\hline
\end{tabular} 
}
\end{table}

%% file: 9-conclusion.tex
\section{Conclusion}
\label{section:Conclusion}

We presented LittleDarwin, a mutation testing framework for Java. On the one hand, it can cope with large and complex  software systems. This lets LittleDarwin foster the adoption of mutation testing in industry.  
On the other hand, the tool is written in Python and released as an open source framework, namely it enables fast prototyping, and the addition of new experimental components. From this point of view, LittleDarwin shows its keen interest in representing an easy to extend framework for researchers on mutation testing.
Combining these aspects allows researchers to use real-life projects as the subjects of their studies.  

In the current version, LittleDarwin is compatible with major build systems, supports complex test structures, can work with large systems, and retains lots of useful information for further analysis of the results.
Moreover, it already includes  the following experimental features: higher-order mutation, mutant sampling, mutant subsumption analysis, and manual mutation. 
Using these features, we have already performed four studies on real-life projects that would otherwise not have been feasible.